\documentclass[aps,pra,showpacs,twocolumn,eqsecnum]{revtex4-2}
\usepackage{graphicx}
\usepackage{amsmath}
\usepackage{amssymb}
\usepackage{amsfonts}
\usepackage{bbm}
\usepackage{bm}
\usepackage{braket}
\usepackage{verbatim}
\usepackage{cancel}
\usepackage{tikz}
\usepackage{wasysym}
\usepackage[bookmarks=false,colorlinks=true,urlcolor=blue,citecolor=blue,linkcolor=blue]{hyperref}
\usepackage[normalem]{ulem}
\allowdisplaybreaks

\tikzstyle{tensor}=[rectangle,draw=blue!50,fill=blue!20,thick]

\begin{document}

\title{Emergent symmetries and slow quantum dynamics in a Rydberg-atom chain with confinement}

\author{I-Chi Chen}
\affiliation{Department of Physics and Astronomy, Iowa State University, Ames, Iowa 50011, USA}
\author{Thomas Iadecola}
\email{iadecola@iastate.edu}
\affiliation{Department of Physics and Astronomy, Iowa State University, Ames, Iowa 50011, USA}

\date{\today}

\begin{abstract}
Rydberg atoms in optical tweezer arrays provide a playground for nonequilibrium quantum many-body physics. The PXP model describes the dynamics of such systems in the strongly interacting Rydberg blockade regime and notably exhibits weakly nonergodic dynamics due to quantum many-body scars. Here, we study the PXP model in a strong staggered external field, which has been proposed to manifest quasiparticle confinement in light of a mapping to a lattice gauge theory. We characterize this confining regime using both numerical exact diagonalization and perturbation theory around the strong-field limit. In addition to the expected emergent symmetry generated by the staggered field, we find a second emergent symmetry that is special to the PXP model. The interplay between these emergent symmetries and the Rydberg blockade constraint dramatically slows down the system's dynamics beyond naive expectations. We devise a nested Schrieffer-Wolff perturbation theory to properly account for the new emergent symmetry and show that this treatment is essential to understand the numerically observed relaxation time scales. We also discuss connections to Hilbert space fragmentation and trace the origin of the new emergent symmetry to a ``nearly-$SU(2)$" algebra discovered in the context of many-body scarring.
\end{abstract}
\maketitle

\section{Introduction}
Rydberg atoms trapped in optical lattices provide a versatile experimental platform to study quantum many-body systems~\cite{Labuhn16,Bernien17,Guardado-Sanchez18,Lienhard18,Ebadi20,Scholl20,Bluvstein21,Morgado20}. Such atomic arrays leverage the strong interactions afforded by exciting neutral atoms to Rydberg states to simulate quantum spin models. In the simplest scenario, each quantum spin is simulated by an individual atom, which can be in either its ground state, $\ket 0$, or a particular Rydberg state, $\ket 1$, so that an effective spin-1/2 model is realized. In the so-called Rydberg blockade regime~\cite{Jaksch00,Urban09}, these interactions are so strong that nearest-neighbor atoms are essentially prohibited from simultaneously occupying the $\ket{1}$ state. Under continuous pumping between the ground and Rydberg states, the effective Hamiltonian for a 1D array of atoms becomes~\cite{Bernien17}
\begin{align}
\label{eq:HPXP}
    H=\lambda\sum_{i}\, P_{i-1}X_iP_{i+1}+\Delta\sum_i n_i,
\end{align}
which is known as the PXP model (see also Ref.~\cite{Fendley04}).
Here  $n_i=(\ket1\bra1)_i$ is the occupation number of the $i$-th atom's Rydberg state, and $X_i=(\ket0\bra1+\ket1\bra0)_i$ flips the atom between its ground and Rydberg states. The parameters $\lambda$ and $\Delta$ correspond to the Rabi frequency of the pump laser and its detuning from the Rydberg state, respectively. The projection operators $P_i=\mathbbm 1-n_i$ enforce the nearest-neighbor (NN) Rydberg blockade constraint.

Despite its simple appearance, the PXP model has a number of intriguing properties. For example, the blockade constraint implies that the model \eqref{eq:HPXP} preserves an exponentially large subspace of Rydberg configurations in which the motif ``11" (representing a pair of NN Rydberg excitations) is forbidden. This constraint reduces the scaling of the Hilbert space for $L$ Rydberg atoms from $2^L$ to $\sim\varphi^L$, where $\varphi=(1+\sqrt{5})/2$ is the golden ratio~\cite{Domb60,Ovchinnikov03,Lesanovsky12b}. For this reason, the NN Rydberg-blockade Hilbert space is sometimes called the Fibonacci Hilbert space. The PXP model is thus an example of a quantum kinetically constrained model, which are known to exhibit slow, glassy dynamics in certain regimes~\cite{VanHorssen15,Pancotti20}.

A related and still-mysterious feature of the PXP model is the existence of quantum many-body scars (QMBS) in the limit of small $\Delta$. QMBS were first observed experimentally in Ref.~\cite{Bernien17}, where the system was prepared in the N\'eel state $\ket{1010\dots}$ and evolved under the Hamiltonian~\eqref{eq:HPXP} with $\Delta=0$. Rather than quickly relaxing to a steady state with no memory of the initial state, as expected from the eigenstate thermalization hypothesis (ETH)~\cite{Deutsch91,Srednicki94,Rigol08,D'Alessio16,Deutsch18}, the system exhibited persistent coherent oscillations of the initial spin pattern. Subsequent theoretical work traced this surprising behavior to a set of ``scarred" eigenstates of the PXP model that have anomalously high overlap with the N\'eel state and are approximately equally spaced in energy~\cite{Turner18a,Turner18b}. While much progress has been made~\cite{Ho19,Khemani18,Lin18,Turner20}, the mechanism underlying these eigenstates of the PXP model is not yet fully understood. However, it is believed that the PXP model is proximate to one in which the scarred eigenstates form an emergent $SU(2)$ multiplet with large total spin~\cite{Choi18,Bull19b}. This ``nearly-$SU(2)$" character also underlies an approximate quasiparticle picture of the scarred eigenstates~\cite{Iadecola19}.

Another remarkable attribute of the PXP model~\eqref{eq:HPXP} is its relationship to $U(1)$ lattice gauge theory. This feature was pointed out and studied in detail in Ref.~\cite{Surace19}, and was foreshadowed by earlier works relating the PXP model to square-lattice quantum dimer models in a quasi-1D limit~\cite{Moessner01,Laumann12,Chen17}. In the gauge theory language, the Rydberg blockade constraint is interpreted as a Gauss's law constraint for a fixed charge background, and the oscillatory dynamics due to QMBS are interpreted as a ``string-inversion" phenomenon.

One consequence of the mapping to a lattice gauge theory is that Eq.~\eqref{eq:HPXP} hosts a regime exhibiting quark-antiquark confinement. This is achieved by replacing the spatially uniform detuning $\Delta$ of Eq.~\eqref{eq:HPXP} with a site-dependent staggered detuning $\Delta_i=(-1)^i(h/2)$. In Ref.~\cite{Surace19}, it was shown that nonzero $h$ induces slow dynamics in both the Rydberg and gauge theory formulations of the model. Although confinement has been investigated in the context of other spin models, e.g. the mixed-field~\cite{McCoy78,Rutkevich08,Kormos17,Pai20} and long-range~\cite{Liu19,Tan19} Ising chains, the PXP model and associated Rydberg-atom platform probes this physics in a different regime. For example, while the aforementioned Ising chains correspond approximately to $\mathbb Z_2$ gauge theories, the PXP model realizes a $U(1)$ Gauss law. Thus, the model provides a route to investigate confinement in continuous gauge theories with quantum simulators. Moreover, the setting of the PXP model begs the question of whether this confining regime bears any relation to QMBS.

In this work, we develop a systematic theory of the confining regime of the PXP model and find a surprising connection to QMBS. In Sec.~\ref{sec:Confinement and Emergent Symmetry}, we use a Schrieffer-Wolff (SW) transformation~\cite{Schrieffer66,Bravyi11} to formulate a perturbative description of confinement in the limit $h\to\infty$, where $h$ is the confining field. This ``strict-confinement" approach was explored in the context of $\mathbb Z_2$ gauge theory in Ref.~\cite{Yang20}, where it was found to yield a hierarchy of time scales that accurately describes the system's dynamics under quenches from initial product states. Surprisingly, we find that this approach fails for the PXP model in the sense that it predicts relaxation times for quench dynamics that do not agree with numerical simulations. In particular, we find a set of initial states whose relaxation times are orders of magnitude longer than the analytical predictions.

In Sec.~\ref{sec:Generalized Schrieffer-Wolff Treatment}, we trace the origin of this failure to the emergence from the SW transformation of Sec.~\ref{sec:Confinement and Emergent Symmetry} of a new approximately conserved quantity. After performing an additional canonical transformation to enforce this new emergent symmetry, we find that the resulting effective Hamiltonian, which we dub a ``nested Schrieffer-Wolff" Hamiltonian, provides an accurate prediction of the numerically observed relaxation time scales. In Sec.~\ref{sec:Hilbert-Space Fragmentation Perspective}, we compare the traditional and nested SW approaches from the perspective of Hilbert-space fragmentation~\cite{Sala19,Khemani19}, which demonstrates why the quench dynamics described by the nested-SW approach are so much slower than those of the traditional approach.

Finally, in Sec.~\ref{sec:SU2} we demonstrate that the emergent conserved quantity that necessitates the nested-SW treatment is a consequence of the nearly-SU(2) algebra first considered in the context of QMBS. Thus, the dramatic slowdown of quench dynamics that we observe numerically and describe analytically using the nested-SW approach is a ramification of this algebra that goes beyond the existence of rare scarred eigenstates and results in a dramatic slowdown of the dynamics of exponentially many initial product states. We also use this algebraic interpretation of our result to identify other Hamiltonians exhibiting slow quantum dynamics.

Our work uncovers a mechanism whereby a many-body system expected to exhibit slow quantum dynamics due to confinement can suffer an additional slowdown due to the unexpected emergence of a new symmetry. In doing so, it also demonstrates a surprising connection between QMBS and confinement in $U(1)$ gauge theory. We highlight possible implications of these findings in our concluding remarks in Sec.~\ref{sec:Conclusion}.

\section{Confinement and Emergent Symmetry}
\label{sec:Confinement and Emergent Symmetry}
We consider a one-dimensional spin-1/2 chain with $L$ lattice sites governed by the Hamiltonian
\begin{align}
\label{eq:Zpi+PXP}
\begin{split}
    H&=h\sum^L_{i=1}(-1)^{i}Z_{i}+\lambda\sum^L_{i=1}P_{i-1}X_{i}P_{i+1}\\
    &\equiv h\, Z_\pi+\lambda\, H_{PXP},
\end{split}
\end{align}
where $Z_i=2n_i-1$ and $X_i$ are Pauli operators on site $i$, and where 
\begin{equation}
P_{i}=\mathbbm 1-n_i=\frac{\mathbbm{1}-Z_{i}}{2}
\end{equation}
is the projector that enforces the Rydberg-blockade constraint.
We work with periodic boundary conditions (PBC) such that $i=L+1\equiv1$ unless specified otherwise.

\begin{figure}[t]
\includegraphics[width=9cm]{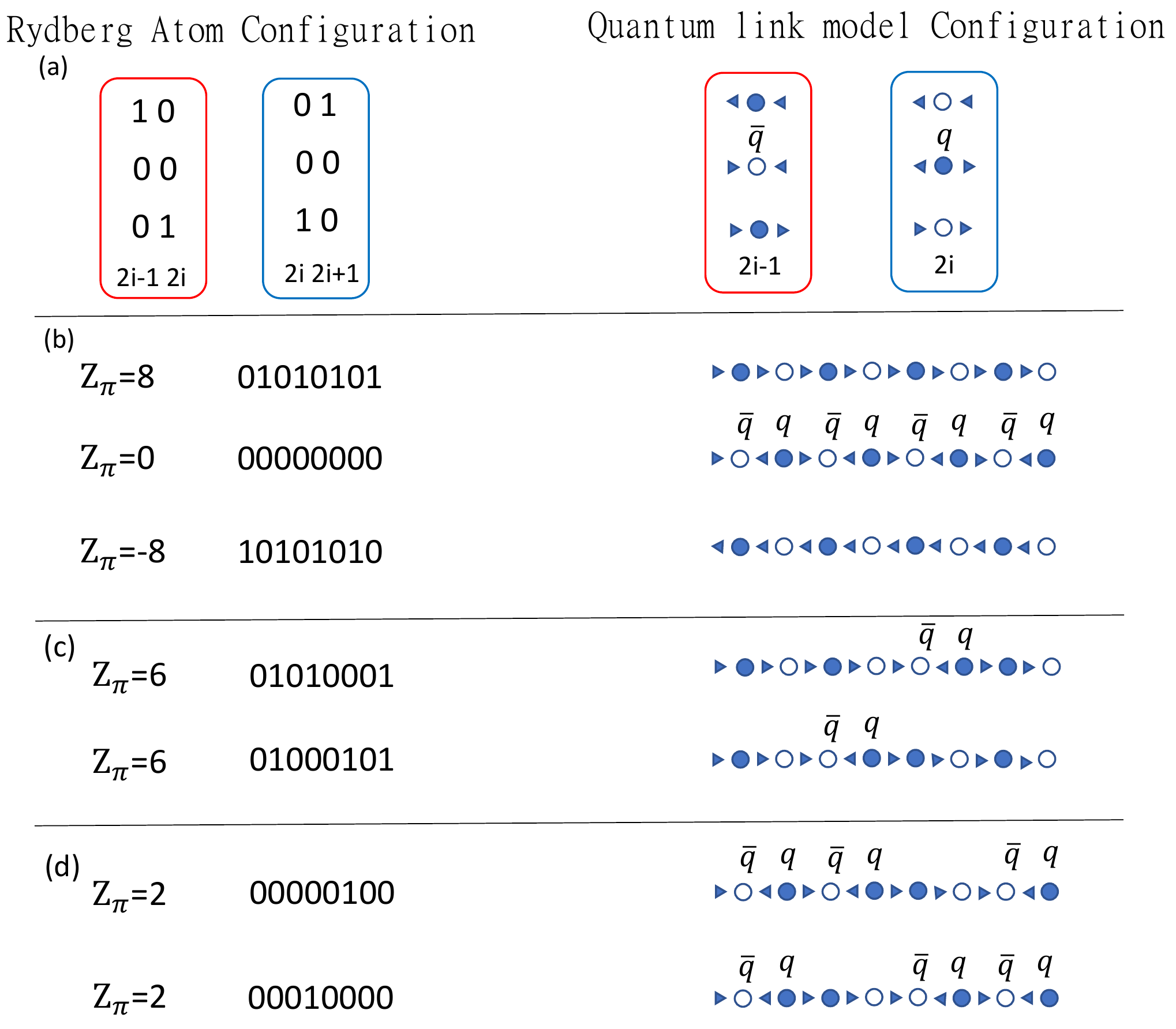}
\centering
\caption{Illustration of the mapping from Rydberg atom to quantum link model configurations introduced in Ref.~\cite{Surace19}. (a) Local mapping of the even-odd and odd-even bonds in the Rydberg chain to even and odd sites, respectively, in the quantum link model. (b) Mapping of the N\'eel and fully polarized states to vacuum and fully packed quark-antiquark states, respectively. Panels (c) and (d) show how the next-nearest-neighbor hopping of Rydberg excitations induced by Eq.~\eqref{eq:Heff4} map to (c) next-nearest-neighbor hopping of quark-antiquark pairs and (d) hopping of a ``hole" in a fully packed quark-antiquark background.}
\label{fig:LGT}
\end{figure}

As shown in Ref.~\cite{Surace19}, $H_{PXP}$ can be mapped onto the spin-$1/2$  quantum link model (QLM), a $U(1)$ lattice gauge theory (LGT), equivalent to a lattice Schwinger model with a topological angle $\theta=\pi$. 
Under this mapping, which is summarized in Fig.~\ref{fig:LGT}(a), Rydberg atom configurations of the form $01$ and $10$ map to a local vacuum configuration of the LGT that contains a gauge field but no particles. The configuration $00$ on sites $2i-1$ and $2i$ maps onto an antiquark in the LGT, while the same configuration on sites $2i$ and $2i+1$ maps onto a quark. With the addition of the staggered field $h\, Z_\pi$, the Rydberg system is equivalent to a lattice Schwinger model with topological angle $\theta\neq\pi$ such that $h\propto \theta-\pi$. In Ref.~\cite{Surace19}, it was shown that a nonzero $h$ leads to confinement of quark-antiquark pairs.

The observed confinement can be interpreted as follows. The operator $Z_\pi$ essentially measures the distance between a quark and an antiquark. Adding this operator to the Hamiltonian with a large coefficient $h$ induces ``string tension," i.e., a large energy cost for separating quarks and antiquarks. In the limit $h\to\infty$, processes that change the distance between quark-antiquark pairs are completely suppressed and $Z_\pi$ becomes an emergent conserved quantity. In this ``strict-confinement" limit the system exhibits Hilbert-space fragmentation, wherein the Hilbert space within a sector with fixed $Z_\pi$ further splits into disconnected subsectors~\cite{Sala19,Khemani19}. This was shown in Ref.~\cite{Yang20} for a $\mathbb Z_2$ gauge theory, and similar phenomenology is expected in more general gauge theories like the $U(1)$ QLM studied here. In this Section we seek to understand the mechanics behind this confinement in further detail. We elaborate on the connection to Hilbert space fragmentation in Sec.~\ref{sec:Hilbert-Space Fragmentation Perspective}.

\subsection{Effective Hamiltonian from Schrieffer-Wolff Transformation}
\label{sec:SW}

We analyze the limit in which the staggered field $h\gg \lambda$.
In this limit, the system's dynamics can be studied perturbatively using an effective Hamiltonian approach based on the Schrieffer-Wolff (SW) transformation~\cite{Schrieffer66,Bravyi11}. In this approach, we rewrite Eq.~\eqref{eq:Zpi+PXP} as
\begin{equation}
    H=H_0+\lambda\, V,
\end{equation}
where $H_0=h\, Z_\pi$ and $V=H_{PXP}$.
The effective Hamiltonian is then defined by performing a unitary transformation generated by an anti-Hermitian operator $S$:
\begin{align}
\label{eq:SW}
    H_{\mathrm{eff}}=e^{S}He^{-S}\equiv H_0+\sum^{\infty}_{n=1} H_{\rm eff}^{(n)},
\end{align}
where $H^{(n)}_{\rm eff}$ is of order $(\lambda/h)^n$.
The Schrieffer-Wolff generator $S$ is chosen such that $[H^{(n)}_{\rm eff},Z_\pi]=0$ for any $n$. This can be accomplished by expanding $S=\sum^{\infty}_{n=1}S^{\left(n\right)}$ with $S^{(n)}$ of order $(\lambda/h)^n$. Substituting this expansion for $S$ into Eq.~\eqref{eq:SW}, expanding $H_{\rm eff}$ using the Baker-Campbell-Hausdorff formula, and collecting terms of order $(\lambda/h)^n$ yields an expression for $H^{(n)}_{\rm eff}$ in terms of $S^{(m)}$ with $m\leq n$.  Then, one can choose $S^{(m)}$ order by order such that $[H_{\rm eff},Z_\pi]=0$~\cite{Bravyi11}. Alternatively, one can choose $S\equiv \lambda \, S^{(1)}$ such that $H^{(1)}_{\rm eff}=0$ and compute higher-order terms using $S$ and projecting onto a fixed eigenspace of $Z_\pi$. We opt for the latter approach here.

Thus, we express
\begin{equation}
\label{eq:BCH}
\begin{split}
    H_{\mathrm{eff}}	&=H_{0}+\lambda\, \mathcal P\left(\left[S^{(1)},H_{0}\right]+V\right)\mathcal P\\
    &\;+\lambda^2\,\mathcal P\left(\left[S^{(1)},V\right]+\frac{1}{2!}\left[S^{(1)},\left[S^{(1)},H_{0}\right]\right]\right)\mathcal P
+\dots,
\end{split}
\end{equation}
where $\mathcal P$ is the projection operator that projects each term in parentheses above onto a fixed eigenspace of $Z_\pi$.
Demanding that $H^{(1)}_{\rm eff}=0$ implies that $S^{(1)}$ must satisfy
\begin{align}
\label{eq:SW-Criterion}
    \left[S^{(1)},H_{0}\right]+V=0.
\end{align}
Substituting this into Eq.~\eqref{eq:BCH} gives the general expression
\begin{align}
\label{eq:Heffn}
    H^{(n)}_{\rm eff}=\lambda^n\, \frac{n-1}{n!}\, \mathcal P\Big[\underbrace{S^{(1)},\Big[S^{(1)},\dots,\Big[S^{(1)}}_{n-1},V\Big]\dots\Big]\Big] \mathcal P
\end{align}
One verifies by direct calculation that Eq.~\eqref{eq:SW-Criterion} is satisfied by the choice
\begin{equation}
\label{eq:S1}
    S^{\left(1\right)}=\frac{i}{2h}\sum_{j}\left(-1\right)^{j}P_{j-1}Y_jP_{j+1}\equiv\frac{i}{2h}H_{PYP}.
\end{equation}
Because $S^{(1)}$ is strictly block off-diagonal in the eigenbasis of $Z_{\pi}$, the odd-order effective Hamiltonians vanish due to the projector $\mathcal P$, while even-order terms survive. 

The leading order effective Hamiltonian is found to be
\begin{equation}
\label{eq:Heff2}
    H^{\left(2\right)}_{\rm eff}=\frac{\lambda^{2}}{2h}\sum_{i}\left(-1\right)^{i}P_{i-1}Z_iP_{i+1}.
\end{equation}
Because this operator is diagonal in the $Z$-basis, it does not induce any transitions between the ground and Rydberg states.
Thus, evolving a $Z$-basis product state with $H^{(2)}_{\rm eff}$ results only in phase accumulation and does not lead to nontrivial dynamics. We will argue in Sec.~\ref{sec:Generalized Schrieffer-Wolff Treatment} that $H^{\left(2\right)}_{\rm eff}$ should be viewed as a second emergent conserved quantity that appears in addition to the expected emergent conservation law for $Z_\pi$ that is enforced by the SW transformation.

The leading-order nontrivial dynamics of this system is therefore generated by the fourth-order effective Hamiltonian. This is given by:
\begin{widetext}
\begin{equation}
\label{eq:Heff4}
\begin{split}
    H_{\mathrm{eff}}^{\left(4\right)}	=\frac{\lambda^{4}}{16h^{3}}\sum_{i}\left(-1\right)^{i}\left[\left(P_{i-1}\sigma_{i}^{+}P_{i+1}\sigma_{i+2}^{-}P_{i+3}
    +
    \text{H.c.}\right)
	+
	P_{i-1}Z_iP_{i+1}P_{i+2}
	+
	P_{i-2}P_{i-1}Z_iP_{i+1}
	-
	2P_{i-1}Z_iP_{i+1}\right],
\end{split}
\end{equation}
\end{widetext}
where $\sigma^\pm_j = (X_j\pm i Y_j)/2$.
The term in parentheses above describes next-nearest-neighbor hopping of Rydberg excitations and implements moves like $010001 \leftrightarrow 000101$. Since the remaining terms are diagonal in the $Z$-basis, the term in parentheses is the one that induces nontrivial dynamics of $Z$-basis product states.
Mapping the action of this term onto the LGT, we find that it produces two kinds of hopping processes. First, it enables the hopping of a nearest-neighbor quark-antiquark pair surrounded by vacuum, as shown in Fig.~\ref{fig:LGT}(c). Second, it enables the hopping of a single ``hole" in the fully-packed quark-antiquark background that corresponds to the polarized state $\ket{0\dots0}$ in the Rydberg system, as shown in Fig.~\ref{fig:LGT}(d). Both of these processes manifestly conserve $Z_\pi$. Moreover, from Eq.~\eqref{eq:Heff4} it is clear that $H_{\mathrm{eff}}^{\left(4\right)}$ preserves the total number of Rydberg excitations, which is related to the total ``magnetization"
\begin{align}
    S_z=\sum_iZ_i.
\end{align} 
This can be viewed as another emergent conserved quantity, in addition to $Z_\pi$. However, unlike $Z_\pi$, $S_z$ is not conserved by higher-order corrections to $H^{(4)}_{\rm eff}$.

\begin{figure}[t]
\includegraphics[width=\columnwidth]{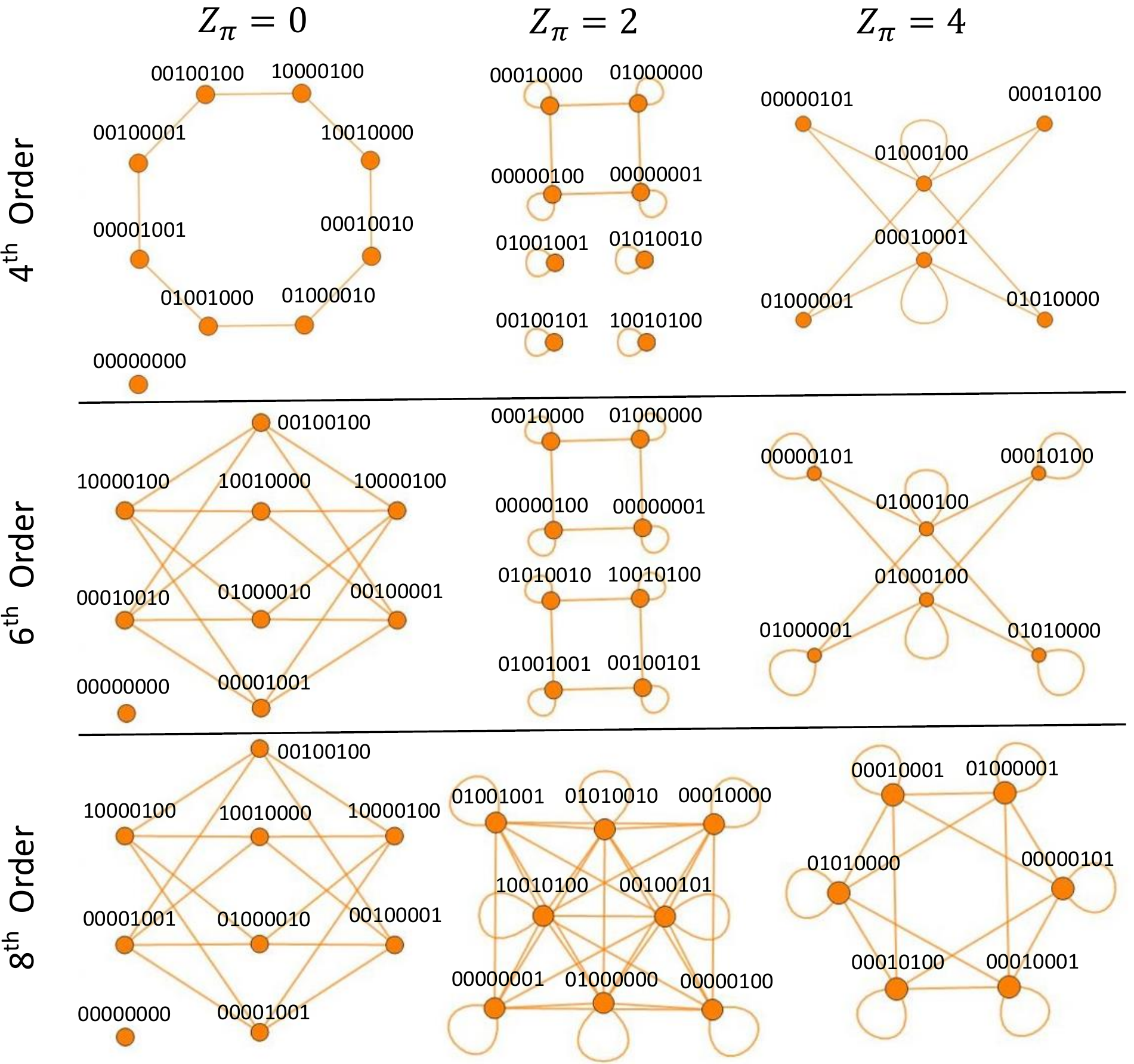}
\centering
\caption{Hilbert space connectivity graph for the SW effective Hamiltonian $H_{\rm eff}$ at $L=8$. The rows correspond to different orders in the SW perturbation theory, while the columns correspond to different values of the emergent conserved quantity $Z_\pi$.}
\label{fig:connectivity}
\end{figure}

Higher order terms in the SW expansion of the effective Hamiltonian become unwieldy to write down explicitly. At a high level, they enable two types of processes. First, they enable further-neighbor hoppings of configurations that are already mobile at fourth order. Second, they enable shorter-range hops of quark-antiquark pairs separated by longer distances. To get a more detailed view of the dynamical processes induced by higher order terms, it is useful to represent these processes graphically as in Fig.~\ref{fig:connectivity}. We represent the Hilbert space within a sector with fixed $Z_\pi$ as a graph whose nodes are $Z$-basis product states and whose edges are the couplings among them.

The first row of Fig.~\ref{fig:connectivity} depicts the fourth-order two-site hopping processes of Rydberg excitations discussed below Eq.~\eqref{eq:Heff4}. 
The middle row of Fig.~\ref{fig:connectivity} depicts processes arising at sixth order, including single-site hoppings of third-neighbor pairs of Rydberg excitations, e.g., $01001\leftrightarrow10010$. Interestingly, we see from the adjacency graph that, like $H^{(4)}_{\rm eff}$, $H^{(6)}_{\rm eff}$ conserves $S_z$. The last row of Fig.~\ref{fig:connectivity} shows processes appearing at 8th order. The adjacency graph in the $Z_\pi=2$ sector demonstrates that $H^{(8)}_{\rm eff}$ no longer conserves $S_z$.

\subsection{Quench Dynamics from Initial Product States}

\begin{figure*}[t]
\includegraphics[width=16cm]{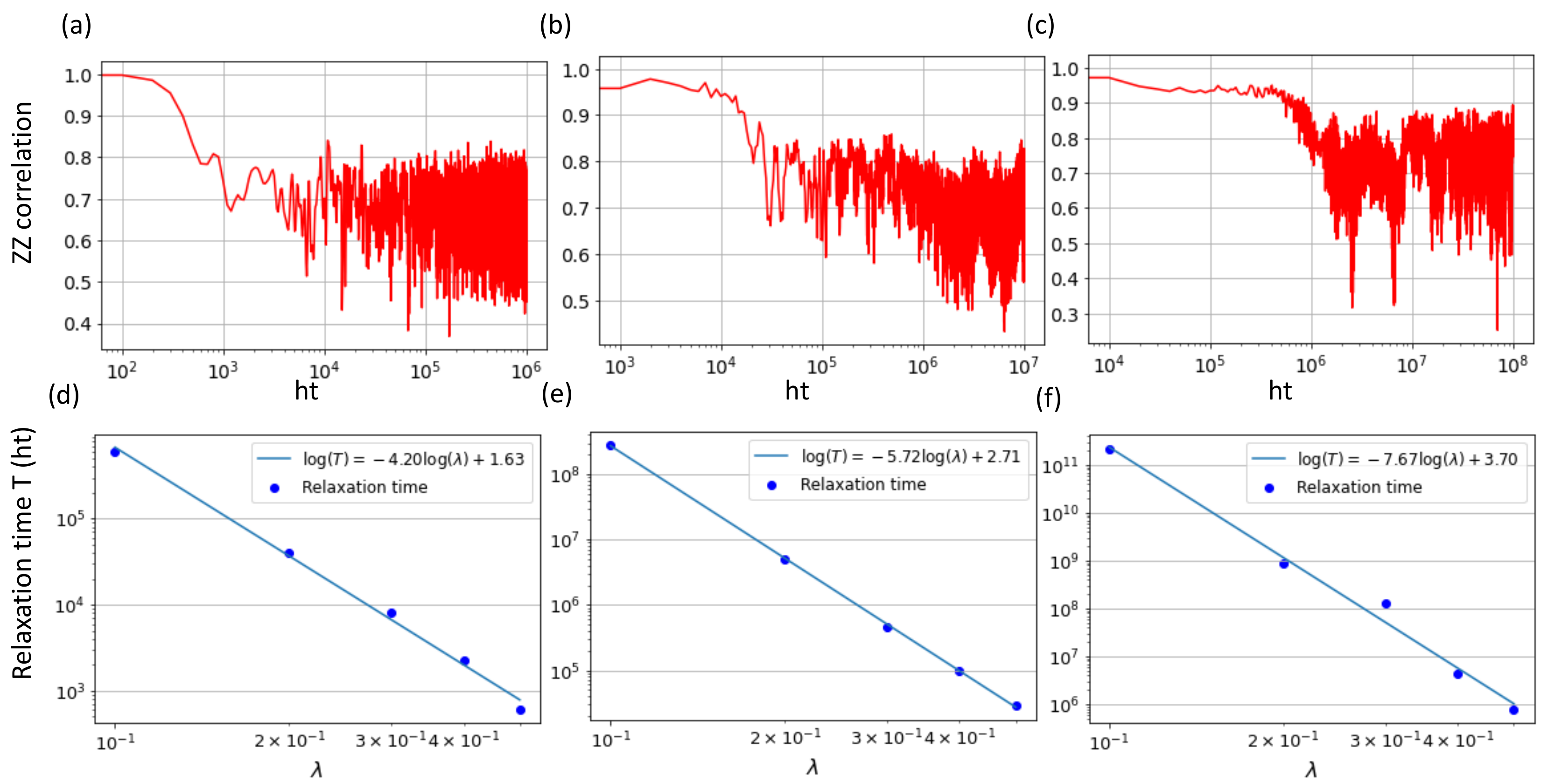}
\centering
\caption{
Dynamics of the autocorrelator $\braket{Z_{L/2}(t)Z_{L/2}(0)}$ with $L=16$. The top row shows the dynamics of the autocorrelator with $\lambda=0.5 h$ for the initial states (a) $\left|\psi_{1}\right\rangle $, (b) $\left|\psi_{2}\right\rangle  $, and (c) $\left|\psi_{3}\right\rangle $ [see Eq.~\eqref{eq:initial states}]. The bottom row shows the dependence of the relaxation time on $\lambda$ for the same initial states, i.e., (d) $\left|\psi_{1}\right\rangle $, (e) $\left|\psi_{2}\right\rangle $, and (f) $\left|\psi_{3}\right\rangle $, on a log-log scale. The solid lines represent power-law fits, whose powers can be read off from the slopes in the insets.
}
\label{fig:ZZ corr}
\end{figure*}

We now test numerically whether the effective Hamiltonian derived in the previous Section captures the system's dynamics. We will restrict our attention to quench dynamics from initial states that are $Z$-basis product states.
When the system is prepared in such a state, the two-time $ZZ$ correlator factorizes, i.e.,
\begin{align}
    \left\langle Z_{i}\left(t\right)Z_{j}\left(0\right)\right\rangle =\left\langle Z_{i}\left(t\right)\right\rangle \left\langle Z_{j}\left(0\right)\right\rangle \propto\left\langle Z_{i}\left(t\right)\right\rangle.
\end{align}
Our investigation of the dynamics will focus on the $ZZ$ autocorrelation function for the spin on site $L/2$, i.e., $\left\langle Z_{L/2}\left(t\right)Z_{L/2}\left(0\right)\right\rangle$.

 In Fig.~\ref{fig:ZZ corr} we show exact diagonalization (ED) results for the quench dynamics of three initial product states at $L=16$, namely
\begin{equation}
\label{eq:initial states}
\begin{split}
     \left|\psi_{1}\right\rangle &=\left|0000000010010000\right\rangle,\\ 
     \left|\psi_{2}\right\rangle &=\left|0000001010010000\right\rangle,\\
     \left|\psi_{3}\right\rangle &=\left|0000101010010000\right\rangle,
\end{split}
\end{equation}
 respectively. In Fig.~\ref{fig:ZZ corr}(a)--(c), we plot the $ZZ$ autocorrelation function $\left\langle Z_{L/2}\left(t\right)Z_{L/2}\left(0\right)\right\rangle$ for $\lambda=0.5\, h$. In all cases, the $ZZ$ autocorrelator remains stable over a few orders of magnitude in time before eventually decaying. 
 
 This slow dynamics can nominally be understood using the effective Hamiltonian picture of Sec.~\ref{sec:SW}. Initially, the Rydberg atom on site $L/2$ in all three product states is in its ground state.
 According to the SW analysis of Sec.~\ref{sec:SW}, the leading-order nontrivial dynamics of the Rydberg excitations in these three states is via next-nearest-neighbor hopping processes described by $H^{(4)}_{\rm eff}$. Such processes enable the following transitions,
\begin{equation}
\begin{split}
\label{eq: 4site}
     \cdots0000010010\cdots&\rightarrow\cdots0100100000\cdots,\\ 
     \cdots000001010010\cdots&\rightarrow\cdots010100100000\cdots,\\
     \cdots00000101010010\cdots&\rightarrow\cdots01010100100000\cdots,
\end{split}
\end{equation}
whereby the configuration of Rydberg excitations in each initial state is translated to the left by four sites.
The result of each of these transitions is to flip the state of the Rydberg atom on site $L/2$ from $0$ to $1$, enabling the relaxation of the $ZZ$ autocorrelator on that site. Since the energy scale of $H^{(4)}_{\rm eff}$ is $\lambda^4/(16h^3)$ [see Eq.~\eqref{eq:Heff4}], we expect that the $ZZ$ autocorrelator for these three initial states will decay at a time of order $t\sim\frac{h^{3}}{\lambda^{4}}$ provided that $\lambda/h$ is sufficiently small.

To test this picture quantitatively, we use ED to calculate the relaxation time of the $ZZ$ autocorrelator as a function of $\lambda$.
Operationally, we define the relaxation time as the first time for which the $ZZ$ autocorrelator is smaller than 0.8.
The results for each initial state are shown in Fig.~\ref{fig:ZZ corr}~(d)--(f). In Fig.~\ref{fig:ZZ corr}~(d), the relaxation time for $\left|\psi_1\right\rangle$ scales as a power law consistent with $t\sim\frac{h^{3}}{\lambda^{4}}$, fitting the naive expectation based on the SW analysis of Sec.~\ref{sec:SW}. However, in Fig.~\ref{fig:ZZ corr}~(b,c), the $ZZ$ autocorrelator of the states $\left|\psi_2\right\rangle $ and $\left|\psi_3\right\rangle $ survives for times orders of magnitude longer than in Fig.~\ref{fig:ZZ corr}~(a), even though the naive expectation based on SW suggests a common $t\sim\frac{h^{3}}{\lambda^{4}}$ scaling for all three initial states. Indeed, Fig.~\ref{fig:ZZ corr} ~(e,f) shows that the relaxation times for $\left|\psi_2\right\rangle $ and $\left|\psi_3\right\rangle $ scale as power laws consistent with $t\sim\frac{h^{5}}{\lambda^{6}}$ and $t\sim\frac{h^{7}}{\lambda^{8}}$, respectively.
This suggests that we need to refine the conventional SW treatment of Sec.~\ref{eq:SW} to understand these dynamics.

\section{Nested Schrieffer-Wolff Treatment}
\label{sec:Generalized Schrieffer-Wolff Treatment}

In this Section we refine the perturbative approach of Sec.~\ref{sec:Confinement and Emergent Symmetry}. The rationale for our refinement is as follows. $H_0 = Z_\pi$ is highly degenerate since there are many $Z$-basis product states with the same $Z_\pi$ eigenvalue.  The SW Hamiltonian describes how degeneracies in the spectrum of $H_0$ are lifted at small perturbation strength $\lambda$. 
However, $H^{(2)}_{\rm eff}$ in Eq.~\eqref{eq:Heff2} is also diagonal in the $Z$-basis. Thus, even though it lifts some degeneracies of $H_0$ (i.e., those arising from $Z$-basis states with the same $Z_\pi$ but different $H^{(2)}_{\rm eff}$), the spectrum of $H^{(2)}_{\rm eff}$ remains highly degenerate. Because the overall energy scale of $H^{(2)}_{\rm eff}$ is much larger than that of $H^{(4)}_{\rm eff}$, we might guess that the operator
\begin{align}
\label{eq:H2}
    H_2=\sum_{i}\left(-1\right)^{i}P_{i-1}Z_iP_{i+1}
\end{align}
should be viewed as another emergent conserved quantity, in addition to $Z_\pi$.  

This guess is consistent with the fact that the fourth-order processes depicted in Eq.~\eqref{eq: 4site} fail to predict the relaxation timescales for the initial states $\ket{\psi_2}$ and $\ket{\psi_3}$ in Fig.~\ref{fig:ZZ corr}. Although the initial and final states in Eq.~\eqref{eq: 4site} all have the same $H_2$ eigenvalue, for $\ket{\psi_2}$ and $\ket{\psi_3}$ the transitions in Eq.~\eqref{eq: 4site} involve intermediate processes like
\begin{equation}
\begin{split}
\label{eq: Heff4 dynamics}
     \cdots0001010010\cdots&\rightarrow\cdots0100010010\cdots,\\ 
     \cdots000101010010\cdots&\rightarrow\cdots010001010010\cdots,
\end{split}
\end{equation}
which do not preserve $H_2$. If the true dynamics of the system are described by an effective Hamiltonian $G_{\rm eff}\neq H_{\rm eff}$ that \textit{does} conserve $H_2$, then the transitions depicted in the second two lines of Eq.~\eqref{eq: 4site} must arise beyond fourth order. This higher-order behavior would result in relaxation times for these initial states that are substantially longer than the ones predicted by the SW analysis of Sec.~\ref{sec:SW}.

\subsection{Derivation of $G_{\rm eff}$}
\label{sec: Geff Derivation}

To derive an effective Hamiltonian $G_{\rm eff}$ that conserves $H_2$, we can perform a second SW transformation $e^{\lambda^{2}S'}$ on $H_{\rm eff}$, i.e.,
\begin{align}
    G_{\rm eff} &= e^{\lambda^{2}S'} H_{\rm eff}e^{-\lambda^{2}S'}\\
&=H_{{\rm eff}}+\lambda^{2}\left[S',H_{{\rm eff}}\right]+\frac{\lambda^{4}}{2}\left[S',\left[S',H_{{\rm eff}}\right]\right]+\ldots,\nonumber
\end{align} 
where the generator $S'$ is an anti-Hermitian operator different from $S$. In order to ensure that $[G_{\rm eff},Z_\pi]=0$, we should demand that $[S',H_0]=0$. By substituting
\begin{align}
H_{\rm eff} = \sum_{n}H^{(n)}_{\rm eff}=\sum_n\lambda^n\tilde{H}^{(n)}_{\rm eff},
\end{align}
where $\tilde{H}^{(n)}_{\rm eff}\equiv H^{(n)}_{\rm eff}/\lambda^n$,
into this expansion and collecting terms of order $\lambda^n$, we can rewrite $G_{\rm eff}$ as
\begin{widetext}
\begin{equation}
\label{eq:Geff}
\begin{split}
G_{{\rm eff}} = &H_{0}+\lambda^{2}\tilde{H}_{{\rm eff}}^{(2)}+\lambda^{4}\left(\tilde{H}_{{\rm eff}}^{(4)}+\left[S',\tilde{H}_{{\rm eff}}^{(2)}\right]\right)+\lambda^{6}\left(\tilde{H}_{{\rm eff}}^{(6)}+\left[S',\tilde{H}_{{\rm eff}}^{(4)}\right]+\frac{1}{2!}\left[S',\left[S',\tilde{H}_{{\rm eff}}^{(2)}\right]\right]\right)+\cdots
\end{split}
\end{equation}
\end{widetext}
We now choose $S'$ such that the term of order $\lambda^4$ conserves $H_2$. To make progress more explicitly, we can break up $\tilde{H}^{(4)}_{\rm eff}$ as
\begin{align}
\label{eq:V4}
\begin{split}
    \tilde{H}^{(4)}_{\rm eff}
    &=\mathcal{P}'\tilde{H}^{(4)}_{\rm eff}\mathcal{P}'+\left(\tilde{H}^{(4)}_{\rm eff}-\mathcal{P}'\tilde{H}^{(4)}_{\rm eff}\mathcal{P}'\right)\\
    &\equiv\mathcal{P}'\tilde{H}^{(4)}_{\rm eff}\mathcal{P}'+V_4.
\end{split}
\end{align}
Here, $\mathcal P'$ is a projector onto a fixed eigenspace of $H_2$, analogous to $\mathcal P$ in Eq.~\eqref{eq:BCH}. Thus, the above decomposition splits $\tilde{H}^{(4)}_{\rm eff}$ into a part that commutes with $H_2$ and another, $V_4$, that does not. Then we can define $S'$ as the solution to the equation
\begin{align}
\label{eq: Generalized SW Criterion}
    \left[\tilde{H}_{{\rm eff}}^{(2)},S'\right]=V_{4},
\end{align}
which removes $V_4$ from the term of order $\lambda^4$ in Eq.~\eqref{eq:Geff}, ensuring that $G^{(4)}_{\rm eff}$ explicitly conserves $H_2$. Note that, since $[V_4,H_0]=0$, Eq.~\eqref{eq: Generalized SW Criterion} also ensures that $[S',H_0]=0$ as desired. Following the analysis of Sec.~\ref{sec:SW}, we substitute Eq.~\eqref{eq: Generalized SW Criterion} into Eq.~\eqref{eq:Geff} and use $\mathcal P'$ to project out the non-$H_2$-conserving parts of the higher-order terms. The new effective Hamiltonian is then
\begin{align}
    G_{{\rm eff}}=H_{0}+\lambda^{2}\tilde{H}_{{\rm eff}}^{(2)}+\lambda^{4}G_{{\rm eff}}^{(4)}+\lambda^{6}G_{{\rm eff}}^{(6)}+\cdots,
\label{eq:Geff1}
\end{align}
where
\begin{subequations}
\begin{align}
    G^{(4)}_{\rm eff}&=\mathcal P'\tilde{H}^{(4)}_{\rm eff}\mathcal P',\\
    \begin{split}
     G_{{\rm eff}}^{(6)}&=\mathcal{P}'\left(\tilde{H}_{{\rm eff}}^{(6)}+\left[S',\tilde{H}_{{\rm eff}}^{(4)}\right]-\frac{1}{2!}\left[S',V_{4}\right]\right)\mathcal{P}'\\
   &=\mathcal{P}'\tilde{H}_{{\rm eff}}^{(6)}\mathcal{P}'+\frac{1}{2!}\mathcal{P}'\left[S',V_{4}\right]\mathcal{P}',
   \label{eq:Geff6}
   \end{split}
\end{align}
\end{subequations}
and so on. The expressions for $\tilde{H}^{(n)}_{\rm eff}$ entering $G_{\rm eff}$ at each order are calculated using Eq.~\eqref{eq:Heffn}. Note that the second equality in Eq.~\eqref{eq:Geff6} above follows from $\mathcal P'[S',P'\tilde{H}^{(4)}_{\rm eff}\mathcal P']\mathcal P'=0$, since $S'$ is block off-diagonal in the eigenbasis of $H_2$ by Eq.~\eqref{eq: Generalized SW Criterion}.

\begin{figure*}[t!]
\includegraphics[width=15cm]{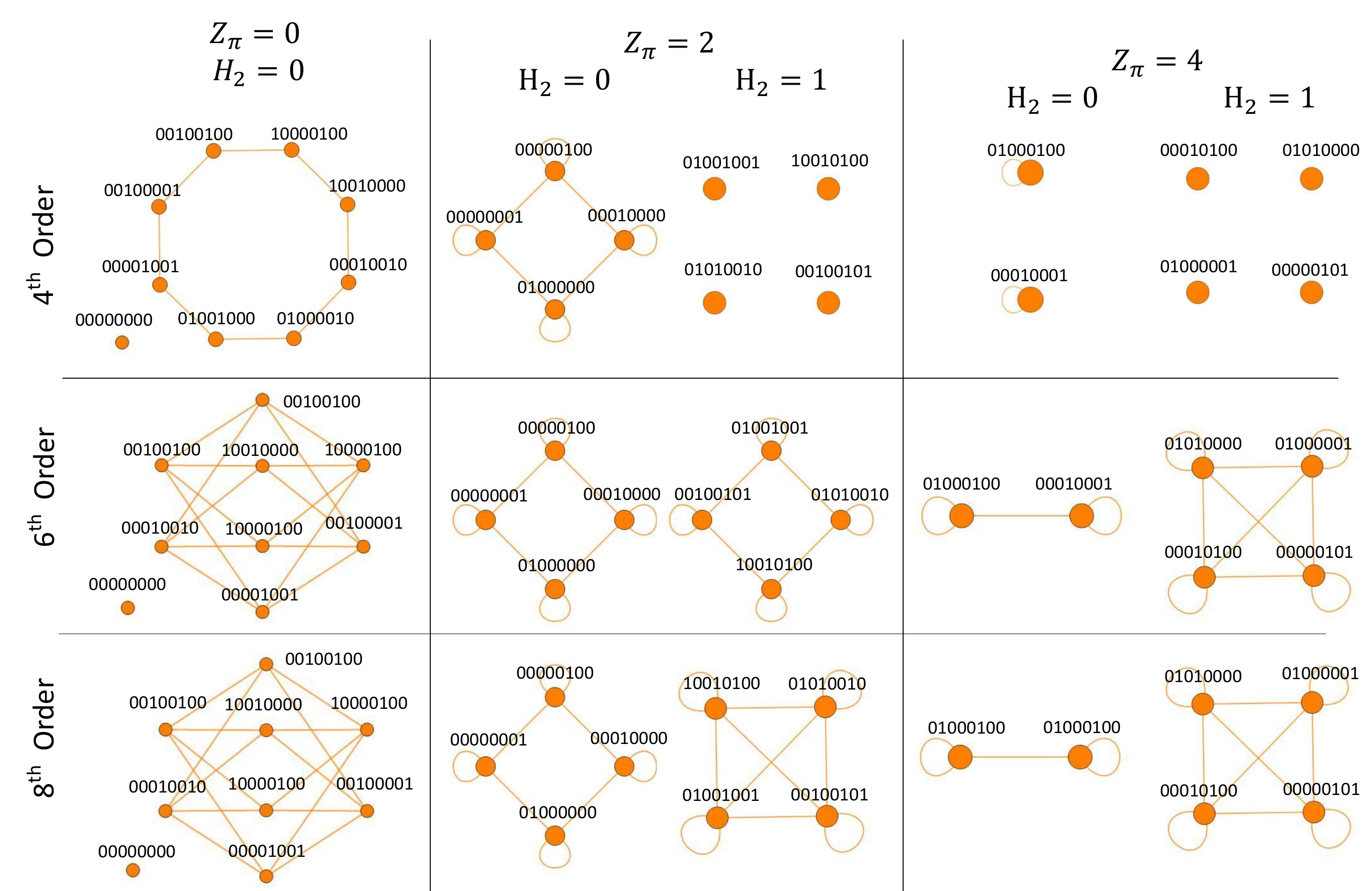}
\centering
\caption{Hilbert space connectivity graph for the nested SW effective Hamiltonian $G_{\rm eff}$. The system size $L=8$, orders in perturbation theory, and $Z_\pi$ sectors considered are the same as in Fig.~\ref{fig:connectivity}, except that now each symmetry sector can be further decomposed into sectors with different $H_2$ eigenvalues.}
\label{fig:General}
\end{figure*}

Calculating the projection of $\tilde{H}^{(4)}_{\rm eff}$, we find that the explicit expression for $G^{(4)}_{\rm eff}$ is
\begin{align}
\label{eq:Geff4}
    G_{{\rm eff}}^{(4)}\propto&\sum_{i}\left(\frac{\mathbbm{1}-Z_{i-2}Z_{i+4}}{2}\right)P_{i-1}\left(\sigma_{i}^{+}P_{i+1}\sigma_{i+2}^{-}+\text{H.c.}\right)P_{i+3} \nonumber \\
    &+\left(\text{diagonal terms}\right).
\end{align}
Evidently the projection further constrains the next-nearest-neighbor hopping terms present in $H^{(4)}_{\rm eff}$, see Eq.~\eqref{eq:Heff4}. For example, transitions like $\cdots0100010\cdots \leftrightarrow \cdots0001010\cdots$ that can occur under $H^{(4)}_{\rm eff}$ are not allowed under $G^{(4)}_{\rm eff}$. One can check explicitly that such processes violate the conservation of $H_2$.

To gain some intuition as to what kinds of processes conserve $H_2$, note that, for any $Z$-basis product state, we have
\begin{align}
    \left\langle H_2\right\rangle = N_{{\rm even}}-N_{{\rm odd}},
\end{align}
where $N_{{\rm even}}$ ($N_{{\rm odd}}$) is the number of three-site ``$101$" N\'eel domains in which the leftmost ``1" resides on an even (odd) site. For instance, the product states $\ket{010100\cdots}$, $\ket{01010100\cdots}$ and $\ket{1010000\cdots}$ have $H_2$ eigenvalues 1, 2 and -1, respectively. 
Translating Rydberg-atom configurations onto LGT configurations using the mapping of Ref.~\cite{Surace19} summarized in Fig.~\ref{fig:LGT}, we find that conservation of $H_2$ amounts to conservation of the number of domain walls between vacuum (i.e., N\'eel) domains and fully packed quark-antiquark (i.e., all-0) domains. 

\begin{figure}[b]
\includegraphics[width=.7\columnwidth]{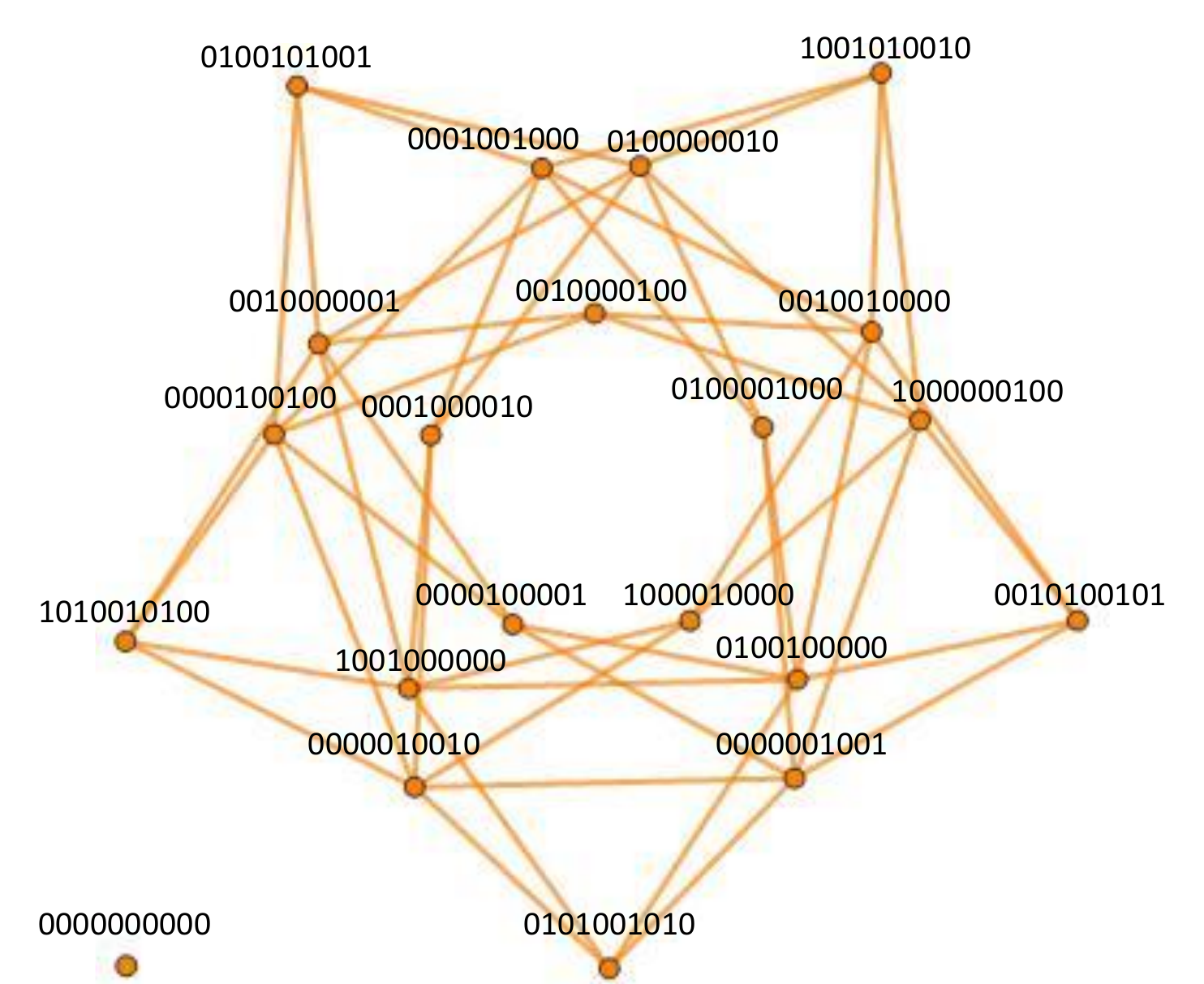}
\centering
\caption{Connectivity graph for $G^{(8)}_{\rm eff}$ with $L=10$ and $Z_{\pi} = 0$. Note that, in contrast with Fig.~\ref{fig:General}, total magnetization $S_z$ is not conserved.}
\label{fig:8th_L10}
\end{figure}

Equipped with the explicit expression for $G^{(4)}_{\rm eff}$ from Eq.~\eqref{eq:Geff4}, one can use Eq.~\eqref{eq:V4} to calculate $V_4$, i.e., the non-$H_2$-conserving part of $\tilde{H}^{(4)}_{\rm eff}$, so that the expressions for higher-order $G^{(n)}_{\rm eff}$ can be derived. However, these expressions are generally unwieldy so we visualize them as connectivity graphs in Fig.~\ref{fig:General} for $L=8$. 

With the added conservation law for $H_2$, the sectors in Fig.~\ref{fig:connectivity} with fixed $Z_\pi$ break up into smaller subsectors, as depicted in the middle and right columns in Fig.~\ref{fig:General}. $H_2$ conservation also restricts the higher order dynamics by preventing ``101" N\'eel domains from splitting into two isolated Rydberg excitations, or two Rydberg excitations from merging into a N\'eel domain. The top-right and middle-right panels of Fig.~\ref{fig:General} show that a ``101" N\'eel domain can move at a time $t\sim\frac{h^{5}}{\lambda^{6}}$ but is frozen at times of order $t\sim\frac{h^{3}}{\lambda^{4}}$ and below. As we discuss in Sec.~\ref{sec: Nested SW dynamics}, this behavior extends to longer N\'eel domains: $H_2$ conservation prevents them  from splitting or merging, so they must remain intact as they move. This implies that motion of a N\'eel domain can only arise at an order in perturbation theory that depends on its length; in particular, our finite-order results suggest that a N\'eel domain of odd length $\ell$ becomes mobile at order $\ell+3$.

From the connectivity graphs in Fig.~\ref{fig:General}, one might be tempted to conclude that the magnetization $S_z$ is also a conserved quantity of $G_{\rm eff}$. However, this is not true at larger system sizes, see, e.g., Fig.~\ref{fig:8th_L10}. For system sizes $L\geq 10$, $S_z$ conservation breaks down at 8th order in the nested SW expansion, similar to what was found in Sec.~\ref{sec:SW}.

\subsection{Dynamics under $G_{\rm eff}$}
\label{sec: Nested SW dynamics}

We now show that the nested SW Hamiltonian $G_{\rm eff}$ can capture the dynamics shown in Fig.~\ref{fig:ZZ corr} for the initial states~\eqref{eq:initial states}. We begin with the state $\ket{\psi_1}$, which consists of the motif ``1001" on a background of 0s. From the leftmost column of Fig.~\ref{fig:General}, we see that the ``1001" motif is already mobile at fourth order, which is consistent with the relaxation timescale $t\sim\frac{h^3}{\lambda^4}$ shown in Fig.~\ref{fig:ZZ corr} (d). The dynamics of this initial state is also consistent with the standard SW treatment of Sec.~\ref{sec:SW}; indeed, the connectivity of the ``1001" motif at each order in Figs.~\ref{fig:connectivity} and \ref{fig:General} is the same.

For $\ket{\psi_{2}}$, the central site $L/2$ is located at the center of a ``101" N\'eel domain. As discussed at the end of Sec.~\ref{sec: Geff Derivation}, this length-3 N\'eel domain does not become mobile until sixth order (see, e.g., the top-right and middle-right panels of Fig.~\ref{fig:General}). Hence, the nested SW treatment suggests that the Rydberg state on site $L/2$ cannot flip until a time $t\sim\frac{h^{5}}{\lambda^{6}}$. This result is consistent with the power law found from numerical results, see Fig.~\ref{fig:ZZ corr}~(e). 

\begin{figure}[t]
\includegraphics[width=\columnwidth]{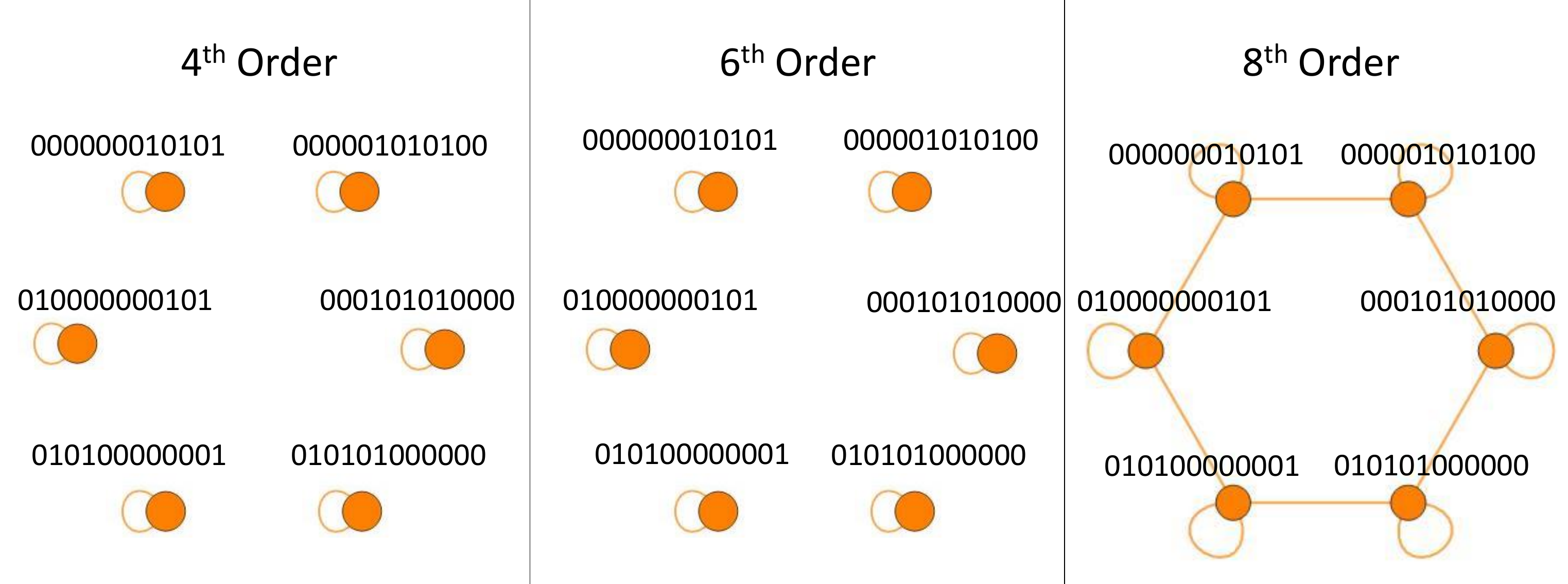}
\centering
\caption{Connectivity graphs for the 4th-, 6th-, and 8th-order nested-SW effective Hamiltonian for $L=12$ and $Z_{\rm \pi} = 6$.}
\label{fig:L12_Zpi_6}
\end{figure}

Finally, for $\ket{\psi_3}$, the central site $L/2$ is contained within a length-5 N\'eel domain ``$10101$." This longer N\'eel domain does not become mobile until eighth order, as we see from the connectivity graph for $L=12$ shown in Fig.~\ref{fig:L12_Zpi_6}. This suggests a relaxation time $t\sim\frac{h^{7}}{\lambda^{8}}$, consistent with the numerical result shown in Fig.~\ref{fig:ZZ corr}~(f). This result also suggests the pattern that a ``101\dots" N\'eel domain of odd length $\ell$ becomes mobile at order $\ell+3$ in the nested SW expansion. 

\section{Hilbert-Space Fragmentation Perspective}
\label{sec:Hilbert-Space Fragmentation Perspective}

In this Section we discuss the connection of our results for the effective Hamiltonians $H_{\rm eff}$ and $G_{\rm eff}$ to earlier works exploring Hilbert space fragmentation (HSF) in gauge theories~\cite{Yang20} and other models~\cite{Sala19,Khemani19,Moudgalya19b,Hudomal19,DeTomasi19,Rakovszky20}. HSF is a phenomenon whereby the Hilbert space of a quantum many-body system ``fractures" into many disconnected components that cannot be uniquely labeled by global symmetry eigenvalues alone. HSF occurs in two varieties, namely ``strong" and ``weak" fragmentation~\cite{Sala19,Khemani19}, that are distinguished by the growth with system size $L$ of the number of disconnected Hilbert-space ``fragments" within a fixed symmetry sector. A system is said to be strongly fragmented if
\begin{align}
\label{eq:Strong HSF}
    \frac{\mathcal D_{\rm max\, subsector}}{\mathcal D_{\rm max\, sector}}\sim e^{-\alpha L},
\end{align}
where $\alpha>0$, $D_{\rm max\, sector}$ is the dimension of the largest symmetry sector of the Hilbert space, and $D_{\rm max\, subsector}$ is the dimension of the largest connected component of that symmetry sector. The above behavior implies a lower bound $\sim e^{\alpha L}$ on the number of Hilbert space ``fragments." A system is said to be weakly fragmented if it is not strongly fragmented.

In Ref.~\cite{Yang20}, it was shown that a $\mathbb Z_2$ gauge theory coupled to fermionic matter in one spatial dimension exhibits HSF in its strict-confinement limit. For the $\mathbb Z_2$ gauge theory, this limit can be treated along the lines of Sec.~\ref{sec:SW}. In particular, the model explicitly conserves fermion number, which in a spin-1/2 representation maps to the Ising domain-wall number
\begin{align}
    n_{\rm DW}=\sum^L_{i=1}\frac{1-Z_iZ_{i+1}}{2}.
\end{align}
In the strict-confinement limit, a SW transformation enforces an additional conservation law for the total magnetization $S_z$. The interplay of these two conservation laws leads to HSF in the effective Hamiltonian $H_{\rm eff}$. In particular, the leading nontrivial order of $H_{\rm eff}$ exhibits strong HSF according to the definition~\eqref{eq:Strong HSF}.

\begin{figure}[t]
\includegraphics[width=\columnwidth]{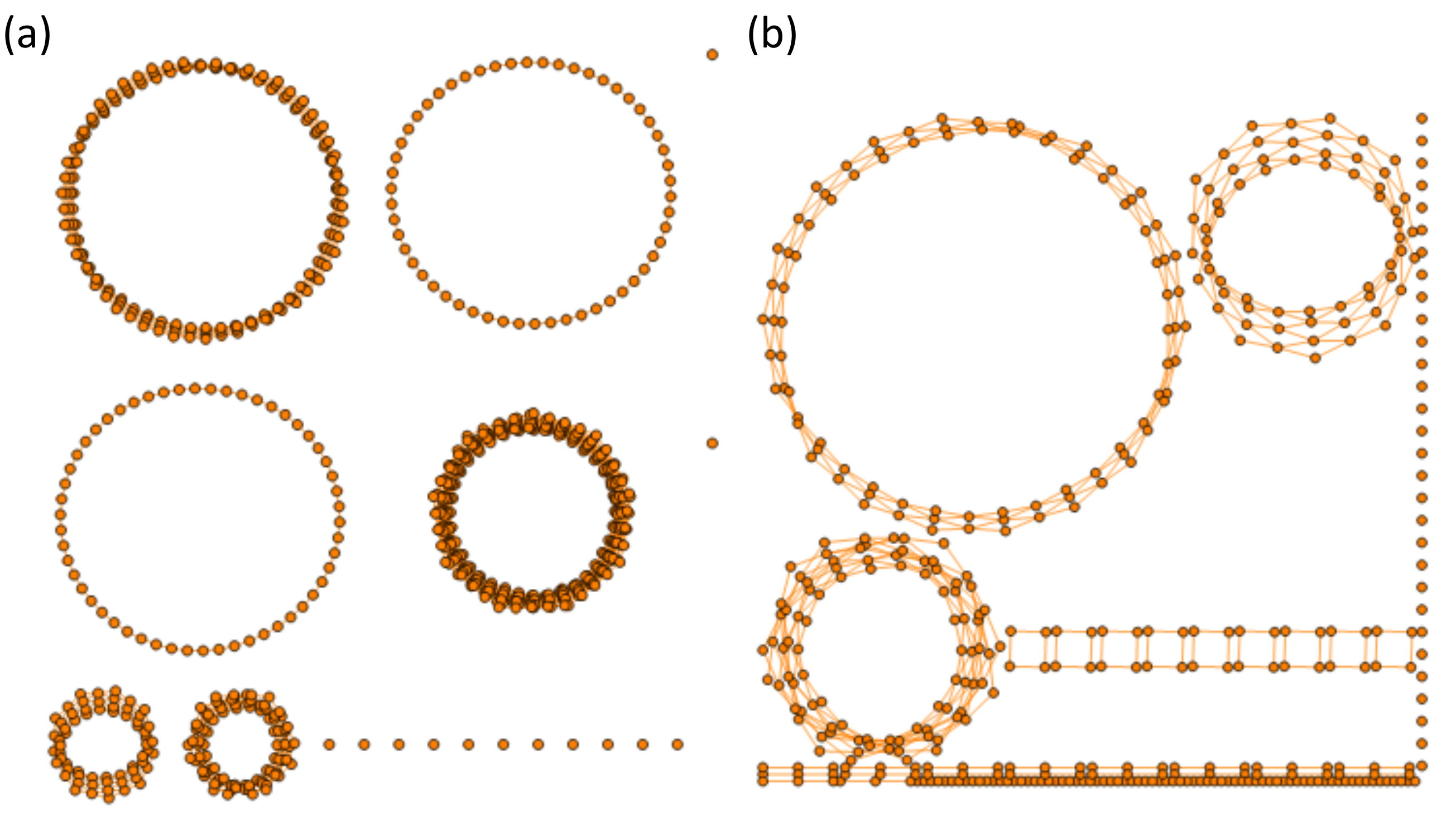}
\centering
\caption{Hilbert space connectivity graphs for (a) the SW effective Hamiltonian $H^{(4)}_{\rm eff}$ at $L=16$ in the sector with $Z_\pi=0$ and (b) the nested SW effective Hamiltonian $G^{(4)}_{\rm eff}$ at $L=16$ in the sector with $Z_\pi=0$ and $H_2=0$.}
\label{fig:H4G4}
\end{figure}

The model \eqref{eq:Zpi+PXP} has a U(1) conservation law associated with the number of nearest-neighbor pairs of Rydberg excitations:
\begin{align}
    n_{\rm NN}=\sum^L_{i=1}n_in_{i+1},
\end{align}
where $n_i=(1+Z_i)/2$ is the Rydberg-state occupation number. The Fibonacci Hilbert space is simply the symmetry sector of states with vanishing $n_{\rm NN}$. In the strict-confinement limit $h\gg \lambda$ of the model \eqref{eq:Zpi+PXP}, an additional U(1) symmetry generated by $Z_\pi$ is imposed. Since $n_{\rm NN}$ is closely related to $n_{\rm DW}$ and $Z_\pi$ is just a staggered version of $S_z$, we thus expect the standard SW Hamiltonian $H_{\rm eff}$ to exhibit strong HSF at leading nontrivial order. The nested SW treatment of Sec.~\ref{sec:Generalized Schrieffer-Wolff Treatment} imposes yet another U(1) symmetry generated by $H_2$, so we expect the nested SW Hamiltonian $G_{\rm eff}$ to be further fragmented at leading order.

\begin{figure}[t]
\includegraphics[width=\columnwidth]{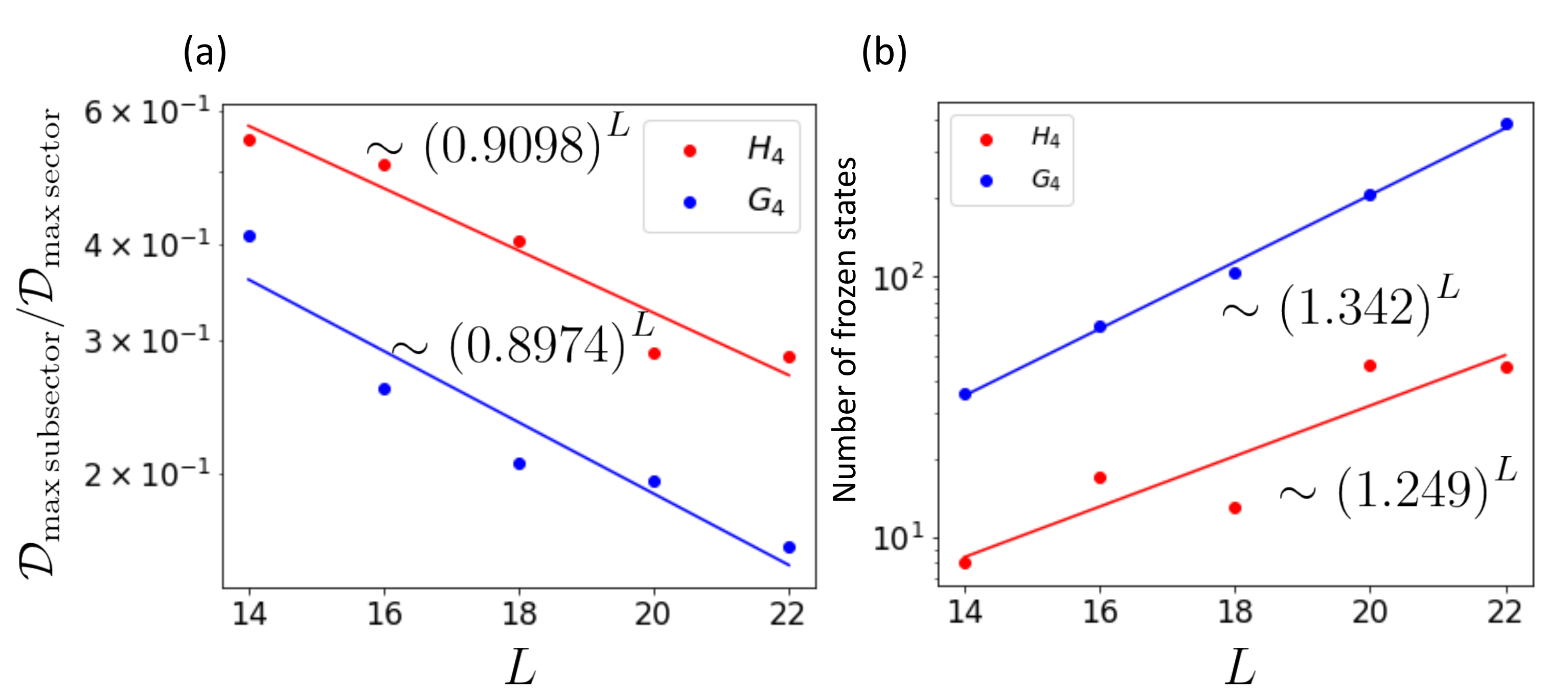}
\centering
\caption{Strong fragmentation and frozen states for the effective Hamiltonians $H^{(4)}_{\rm eff}$ and $G^{(4)}_{\rm eff}$. (a) Scaling with $L$ of the strong-HSF diagnostic \eqref{eq:Strong HSF}, suggesting that both $H^{(4)}_{\rm eff}$ (red) and $G^{(4)}_{\rm eff}$ (blue) exhibit strong HSF. (b) Scaling of the number of frozen states (i.e., connected components with dimension 1) with system size for $H^{(4)}_{\rm eff}$ (red) and $G^{(4)}_{\rm eff}$ (blue). Both models exhibit an exponentially growing number of frozen states, with $G^{(4)}_{\rm eff}$ having exponentially more such states than $H^{(4)}_{\rm eff}$. In both (a) and (b), the symmetry sectors used to calculations involving $H^{(4)}_{\rm eff}$ and $G^{(4)}_{\rm eff}$ match those used in Fig.~\ref{fig:H4G4}.}
\label{fig:Fragment-Scaling}
\end{figure}

To test this hypothesis numerically, we plot in Fig.~\ref{fig:H4G4} (a) and (b) the Hilbert space connectivity graphs for the leading nontrivial effective Hamiltonians $H^{(4)}_{\rm eff}$ and $G^{(4)}_{\rm eff}$, respectively, at $L=16$. In both cases, we focus on the largest symmetry sector, namely the subsector of the Fibonacci Hilbert space with $Z_\pi=0$ ($Z_\pi=H_2=0$) for the (nested) SW Hamiltonian $H^{(4)}_{\rm eff}$ ($G^{(4)}_{\rm eff}$). Fig.~\ref{fig:H4G4} provides qualitative validation of our hypothesis: in both (a) and (b) the connectivity graph breaks into many disconnected subsectors, including many that contain a single product state. (Product states that constitute their own connected component of the Hilbert space are called ``frozen states.")  The $G^{(4)}_{\rm eff}$ connectivity graph in Fig.~\ref{fig:H4G4} (b) has markedly more disconnected subsectors despite the full Hilbert space containing fewer states due to the additional constraint of $H_2$ conservation.

Fig.~\ref{fig:Fragment-Scaling} provides a quantitative comparison of fragmentation in the models $H^{(4)}_{\rm eff}$ and $G^{(4)}_{\rm eff}$, focusing on the same symmetry sectors as in Fig.~\ref{fig:H4G4}. In Fig.~\ref{fig:Fragment-Scaling}~(a) we calculate the fragmentation diagnostic~\eqref{eq:Strong HSF} as a function of $L$ and find exponential decay for both $H^{(4)}_{\rm eff}$ and $G^{(4)}_{\rm eff}$, indicating that both models indeed exhibit strong HSF. The diagnostic \eqref{eq:Strong HSF} appears to decay slightly faster with $L$ for $G^{(4)}_{\rm eff}$, indicating that the lower bound on the number of ``fragments" grows faster for $G^{(4)}_{\rm eff}$ than it does for $H^{(4)}_{\rm eff}$. This is consistent with Fig.~\ref{fig:H4G4}, which shows that the Hilbert space of the model $G^{(4)}_{\rm eff}$ is indeed ``more fragmented" than that of $H^{(4)}_{\rm eff}$ at fixed $L$. 

In Fig.~\ref{fig:Fragment-Scaling}~(b) we calculate the scaling with $L$ of the number of frozen states (i.e., connected components containing a single product state) for the two models. It is clear that $G^{(4)}_{\rm eff}$ has substantially more frozen states than $H^{(4)}_{\rm eff}$, despite the fact that the former model has a smaller Hilbert space than the latter due to the additional conserved quantity $H_2$. 

This difference in the scaling of the number of frozen states is the origin of the discrepancy in dynamical time scales between the SW treatment of Sec.~\ref{sec:SW} and the nested-SW treatment of Sec.~\ref{sec:Generalized Schrieffer-Wolff Treatment}. Many initial product states that have nontrivial dynamics at fourth order in the SW treatment described by $H_{\rm eff}$ are in fact frozen in the nested-SW treatment described by $G_{\rm eff}$. These initial states do not obtain nontrivial dynamics until higher orders in the nested-SW approach, which introduce longer-range terms that reduce the degree of HSF and with it the number of frozen states.
 
\section{Relation to Nearly-$SU(2)$ Algebra}
\label{sec:SU2}
We now connect the slow dynamics studied in this paper to the phenomenon of QMBS in the PXP model. In the process, we will learn more about the stability of these slow dynamics to the addition of other types of Hamiltonian terms.

The main reason underlying the slowness of the dynamics under the Hamiltonian~\eqref{eq:Zpi+PXP} is the fact that $H^{(2)}_{\rm eff}$ is diagonal in the $Z$ basis. This both delays the leading-order nontrivial dynamics to order $\lambda^4$ and generates the additional emergent symmetry $H_2$ [Eq.~\eqref{eq:H2}] that further fragments the Hilbert space at fixed orders in perturbation theory. The diagonality of $H^{(2)}_{\rm eff}$ is a consequence of the fact that $H_0$ and $V$ in Eq.\eqref{eq:Zpi+PXP} can be viewed as generators of a ``nearly-$SU(2)$" algebra. This algebra was studied in the context of QMBS in the PXP model in Refs.~\cite{Choi18,Iadecola19,Bull19b}. Here, we demonstrate that it gives rise to deformations of Eq.~\eqref{eq:Zpi+PXP} that also yield slow dynamics. 

The nearly-$SU(2)$ algebra is generated by the operators $Z_\pi$, $H_{PXP}$, and $H_{PYP}$, where the latter is defined in Eq.~\eqref{eq:S1}. For PBC, the commutation relations among these generators are given by 
\begin{equation}
\begin{split}
\label{eq: Nearly SU2}
    \left[Z_{\pi},H_{PXP}\right]&=2i\,H_{PYP}\\
    \left[Z_{\pi},H_{PYP}\right]&=-2i\,H_{PXP}\\
    \left[H_{PXP},H_{PYP}\right]&=2i\left(Z_{\pi}+\mathcal{O}_{zzz}\right)= 2i\, H_2,
\end{split}
\end{equation}
where $\mathcal{O}_{zzz}=\sum_{j}\left(-1\right)^{j}Z_{j-1}Z_{j}Z_{j+1}$. From the above, we see that $H_0$ and $V$ in Eq.~\eqref{eq:Zpi+PXP} are the ``$Z$" and ``$X$" generators of the nearly-$SU(2)$ algebra. The SW generator $S^{(1)}$ [Eq.~\eqref{eq:S1}] that eliminates off-diagonal matrix elements of $H$ to leading order is related to the ``$Y$" generator of this algebra. If the algebra \eqref{eq: Nearly SU2} were closed, we would have $H^{(2)}_{\rm eff}\propto Z_\pi$. Instead, we obtain $H^{(2)}_{\rm eff}\propto H_2$, which shares an eigenbasis with $Z_\pi$. (We note in passing that this would not occur if we had replaced $Z_\pi$ by the non-staggered magnetization $S_z$ and then carried out the SW transformation.) This connection motivates us to consider Hamiltonians consisting of other combinations of nearly-$SU(2)$ generators.

First, we consider adding a term proportional to $H_{PYP}$ to the Hamiltonian \eqref{eq:Zpi+PXP}, i.e.,
\begin{align}
\label{eq:ZXY Hamiltonian}
     H=h\, Z_{\pi}+\lambda_1\, H_{PXP}+\lambda_2\, H_{PYP}.
\end{align}
Following the analysis of Sec.~\ref{sec:SW}, we consider the limit $h\gg \lambda_1, \lambda_2$ and find the generator $S^{(1)}$ analogous to Eq.~\eqref{eq:S1} using the commutation relations in Eq.~\eqref{eq: Nearly SU2}:
\begin{align}
S^{(1)} \propto \lambda_1\, H_{PYP}-\lambda_2\, H_{PXP}.
\end{align}
From Eq.~\eqref{eq:Heffn}, one finds the second-order effective Hamiltonian
\begin{align}
\begin{split}
H_{{\rm eff}}^{(2)}&\propto\left(\lambda_1^2+\lambda_2^2\right)\left[H_{PYP},H_{PXP}\right]\\
&\propto\left(\lambda_1^2+\lambda_2^2\right)H_2,
\end{split}
\end{align}
similar to Eq.~\eqref{eq:Heff2}. It follows that the slow dynamics and nested-SW approach studied in this paper apply equally well to the Hamiltonian~\eqref{eq:ZXY Hamiltonian}.

\begin{figure}[t]
\includegraphics[width=\columnwidth]{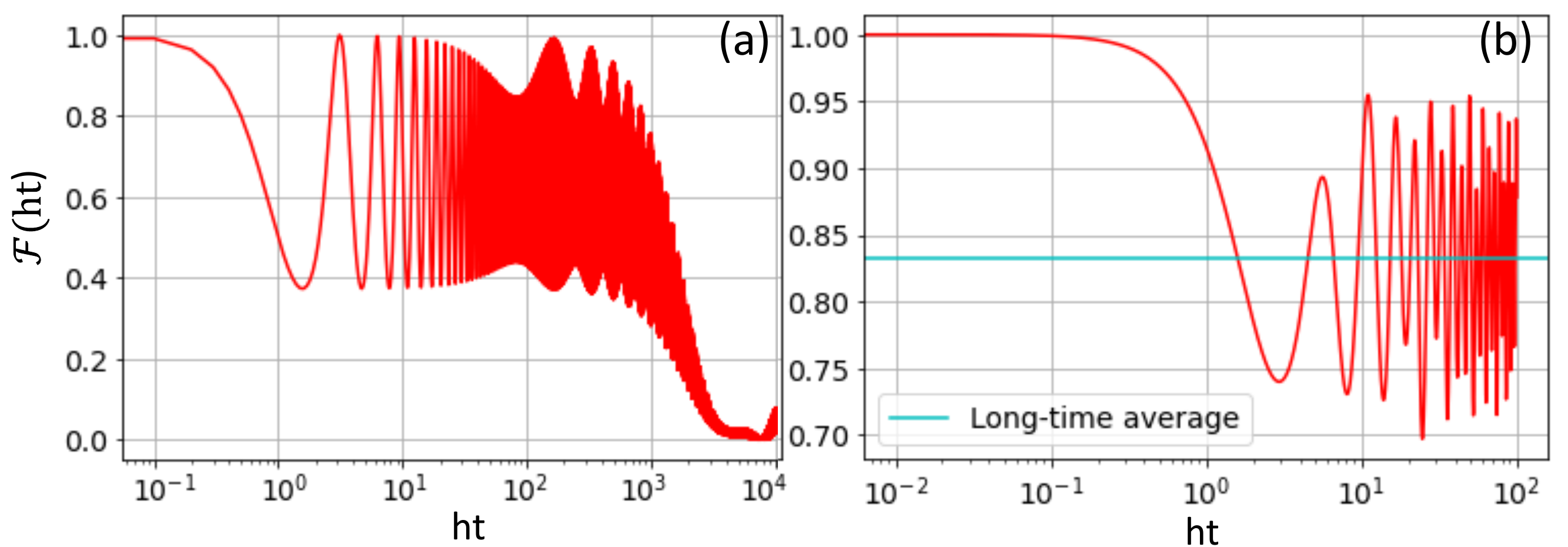}
\centering
\caption{Fidelity dynamics for the Hamiltonians of (a) Eq.~\eqref{eq:ZXY Hamiltonian} and (b) Eq.~\eqref{eq:PXP+PYP} at $L=16$. The Hamiltonian parameters in (a) are $\lambda_1=\lambda_2=0.2 h$, and the initial state is chosen to be $\ket{\psi_1}$ [see Eq.~\eqref{eq:initial states}]. The Hamiltonian parameters in (b) are $\lambda=0.2\, h$ and the initial state is chosen to be the $325$th eigenstate of $H_{\rm PXP}$, which is nondegenerate. The horizontal line indicates the late-time value $\mathcal F_\infty=0.83\dots$ calculated from Eq.\eqref{eq:F_inf}.}
\label{fig:Fid}
\end{figure}

Another Hamiltonian amenable to the above analysis is 
\begin{align}
\label{eq:PXP+PYP}
     H=h\, H_{PXP}+\lambda\, H_{PYP}.
\end{align}
In the limit $h\gg \lambda$, the SW generator can be calculated as
\begin{align}
\label{eq:PXP+PYP S1}
     S^{(1)}\propto Z_\pi,
\end{align}
and the corresponding second-order effective Hamiltonian is given by
\begin{align}
H_{{\rm eff}}^{(2)}\propto\left[H_{PYP},Z_\pi\right] \propto H_{PXP}.
\end{align}
This suggests a slowdown of quench dynamics when the system is prepared in an eigenstate of $H_{PXP}$. We remark that the above analysis also holds with $H_{PXP}$ and $H_{PYP}$ interchanged. However, it breaks down if a term $\lambda_2 Z_\pi$ is added due to the fact that the algebra~\eqref{eq: Nearly SU2} is not closed.

We now test numerically whether the Hamiltonians \eqref{eq:ZXY Hamiltonian} and \eqref{eq:PXP+PYP} inspired by the nearly-$SU(2)$ algebra exhibit slow dynamics. In Fig.~\ref{fig:Fid}, we plot ED results at $L=16$ for the many-body fidelity, defined as
\begin{align}
\mathcal F(t)=\left|\braket{ \psi(t)|\psi(0)} \right|^{2}.
\end{align}
In Fig.~\ref{fig:Fid} (a), we show fidelity dynamics for the Hamiltonian \eqref{eq:ZXY Hamiltonian} with $\lambda_1=\lambda_2=0.2\, h$ with the the initial state $\ket{\psi_1}$ [see Eq.~\eqref{eq:initial states}]. A relaxation timescale $t\sim\frac{h^{3}}{\lambda^{4}}$ is observed, in agreement with the nested SW description of Sec.~\ref{sec:Generalized Schrieffer-Wolff Treatment}. In Fig.~\ref{fig:Fid} (b), we show fidelity dynamics for the Hamiltonian \eqref{eq:PXP+PYP} for $\lambda=0.2\, h$ with the initial state taken to be the $325$th eigenstate of $H_{PXP}$, which is nondegenerate. In this case, the fidelity quickly relaxes and oscillates around a finite average value that persists to all numerically accessible times. The fact that the fidelity remains finite to arbitrarily late times is highly atypical---for example, even in Fig.~\ref{fig:Fid} (a), the fidelity eventually relaxes to a value close to zero.

In fact, the finite value of the fidelity observed in Fig.~\ref{fig:Fid} (b) has a nonperturbative explanation: the Hamiltonian~\eqref{eq:PXP+PYP} is unitarily equivalent to $H_{PXP}$. To see this, note that the algebra~\eqref{eq: Nearly SU2} implies that the unitary transformation
\begin{subequations}
\begin{align}
    U(\theta) = e^{-i\theta Z_\pi/2}
\end{align}
acts on the Hamiltonian~\eqref{eq:PXP+PYP} as
\begin{align}
    U^\dagger(\theta)\, H\, U(\theta)=\sqrt{h^2+\lambda^2}\, H_{PXP},
\end{align}
\end{subequations}
when $\theta = \tan^{-1}(\lambda/h)$. Consequently, any eigenstate $H_{PXP}\ket{\psi_E}=E\ket{\psi_E}$ of $H_{PXP}$ corresponds to an eigenstate $U(\theta)\ket{\psi_E}$ of $H$ with the same energy $E$, up to a rescaling by $\sqrt{h^2+\lambda^2}$. 
Using this fact, we can express the long-time average of $\mathcal F(t)$ for the initial state $\ket{\psi_E}$ as
\begin{align}
\label{eq:F_inf}
    \mathcal F_\infty\equiv \overline{\mathcal F(t\to\infty)} = \sum_{E'}|\braket{\psi_E|U(\theta)|\psi_{E'}}|^4,
\end{align}
which is the inverse participation ratio of the initial state $\ket{\psi_E}$ in the eigenbasis $\{U(\theta)\ket{\psi_{E'}}\}$ of $H$.
When $\lambda \ll h$, $\theta\approx \lambda/h\ll 1$, so $U(\theta)$ is close to the identity operator and the overlap $\braket{\psi_E|U(\theta)|\psi_E}$ is finite, leading to a finite value of $\mathcal F_{\infty}$. An example of this is shown in Fig.~\ref{fig:Fid} (b), where the horizontal line denotes the value of $\mathcal F_\infty$ calculated from the above expression. If instead the eigenbases of $H$ and $H_{PXP}$ were unrelated, we would expect $\mathcal F_{\infty}$ to be exponentially small; this is the case for quantum quenches between generic many-body Hamiltonians~\cite{Gorin06,Goussev16}.

Thus, our results in this Section show that the nearly-$SU(2)$ algebra \eqref{eq: Nearly SU2} can give rise to two types of atypical quantum quench dynamics. The first type, exemplified by the Hamiltonian \eqref{eq:ZXY Hamiltonian}, is the slow dynamics studied in Secs.~\ref{sec:Confinement and Emergent Symmetry}--\ref{sec:Hilbert-Space Fragmentation Perspective}, which is governed by the pair of emergent symmetries $Z_\pi$ and $H_2$. In this type of dynamics, a hierarchy of relaxation time scales follows from perturbation theory in the limit of large staggered field. The second type of dynamics, exemplified by the Hamiltonian \eqref{eq:PXP+PYP}, is characterized by a finite late-time average of the fidelity. The fact that the first type of dynamics arises at all is a consequence of the fact that the algebra~\eqref{eq: Nearly SU2} is not closed. If instead Eq.~\eqref{eq: Nearly SU2} were a genuine $SU(2)$ algebra, one could always perform an appropriate $SU(2)$ rotation to bring any Hamiltonian of the form \eqref{eq:ZXY Hamiltonian} to one proportional to any of the generators $Z_\pi,H_{PXP}$, or $H_{PYP}$.

\section{Conclusion}
\label{sec:Conclusion}

In this work, we have developed a theory of the confining regime of the PXP model, which was proposed in Ref.~\cite{Surace19} based on a connection to $U(1)$ lattice gauge theory. Intriguingly, we find that an analysis of the confining regime along the lines of Ref.~\cite{Yang20}, which considered a model relevant to $\mathbb Z_2$ lattice gauge theory, is not sufficient to describe the hierarchy of relaxation time scales observed in numerical simulations of quench dynamics. Instead we show that the PXP model, with the addition of a large staggered longitudinal field $Z_\pi$, exhibits a new emergent conserved quantity $H_2$ that further constrains the dynamics. We devise a nested Schrieffer-Wolff approach that properly accounts for this emergent symmetry and show that this approach explains the numerically observed dynamics. We also verify that the emergent symmetry $H_2$ leads, at leading nontrivial order in the effective Hamiltonian description, to a substantially more fragmented Hilbert space relative to what one finds using the approach of Ref.~\cite{Yang20}. Finally, we trace the origin of the emergent conserved quantity $H_2$ to the nearly-$SU(2)$ algebra believed to underlie many-body scarring in the PXP model at zero staggered field. This surprising algebraic connection to scars in fact underlies the inability of the ``standard" Schrieffer-Wolff approach of Ref.~\cite{Yang20} to describe the dynamics in this case.

One potentially fruitful direction for future work is to explore the existence of algebraically induced slow dynamics in higher-dimensional generalizations of the PXP model~\cite{Celi20,Lin20,Michailidis20}, which are also experimentally realizable~\cite{Ebadi20}. For example, a nearly-$SU(2)$ algebra has also been uncovered in the 2D PXP model on the square lattice~\cite{Michailidis20}. It is therefore natural to expect a regime of slow dynamics in that model that can be treated using the nested-SW approach developed here. 

Another possible direction for future work is to clarify the role of the emergent conserved quantity $H_2$ in the $U(1)$ lattice gauge theory formulation of the model. In the gauge theory, the staggered magnetization $Z_\pi$ is related to the detuning of the topological $\theta$-angle from $\pi$. It would be interesting to determine how $H_2$ fits into this description. For example, the algebraic relationship between $Z_\pi$ and $H_2$ [see Eq.~\eqref{eq: Nearly SU2}] seems to suggest that emergent $H_2$ conservation is related to the continuous nature of the gauge group $U(1)$. This relationship, if it exists, may explain why the techniques used in Ref.~\cite{Yang20} to treat a discrete gauge theory are not sufficient to describe the dynamics in the $U(1)$ case. A related direction is to consider whether similar phenomena arise in quantum-link-model formulations of non-Abelian gauge theories~\cite{Orland90,Chandrasekharan97,Banerjee13}.

\begin{acknowledgments}
We thank Alexey Gorshkov, Fangli Liu, and Zhicheng Yang for valuable discussions and collaboration on related work. This work was supported by Iowa State University startup funds.
\end{acknowledgments}

\bibliography{refs}

\end{document}